# Realisation of the first sub shot noise wide field microscope

Nigam Samantaray[1,2], Ivano Ruo-Berchera[1], Alice Meda[1,*], Marco Genovese[1,3]

[1] *INRIM, Strada delle Cacce 91, I-10135 Torino, Italy*

[2] *Politecnico di Torino, Corso Duca degli Abruzzi, 24 - I-10129 Torino, Italy and*

[3] *INFN, Via P. Giuria 1, I-10125 Torino, Italy*[*]

[*] Correspondence: Alice Meda, Email: a.meda@inrim.it, Tel: +390113919245

**In the last years several proof of principle experiments have demonstrated the advantages of quantum technologies respect to classical schemes. The present challenge is to overpass the limits of proof of principle demonstrations to approach real applications. This letter presents such an achievement in the field of quantum enhanced imaging. In particular, we describe the realization of a sub-shot noise wide field microscope based on spatially multi-mode non-classical photon number correlations in twin beams. The microscope produces real time images of 8000 pixels at full resolution, for $(500 \mu m)^2$ field-of-view, with noise reduced to the 80% of the shot noise level (for each pixel), suitable for absorption imaging of complex structures. By fast post-elaboration, specifically applying a quantum enhanced median filter, the noise can be further reduced (less than 30% of the shot noise level) by setting a trade-off with the resolution, demonstrating the best sensitivity per incident photon ever achieved in absorption microscopy.**

**INTRODUCTION**

Sensitivity in standard optical imaging and sensing, the ones exploiting classical illuminating fields, is fundamentally lower bounded by the shot noise, the inverse square root of the number of photons used. Beating such a limit is particularly effective when there is a constrain on the usable optical power, for example determined by the damage threshold of the sample[1], the stress of the optical elements[2], or alteration of chemical and biological photo sensitive process and, most fundamentally, quantum back-action[3]. Sprouting from the seminal works of Caves[4], showing how squeezed light could improve the sensitivity in interferometry, non-classical states of light have been considered for a long time to overcome shot noise, giving rise to a deep theoretical investigation and many proposed schemes[5-9].

The experimental possibility to generate two-photon entangled states[10] (such as NOON states with N=2) and the availability of single photon detectors have enabled the demonstration of the potentiality of quantum enhanced sensing, aimed at reaching the fundamental Heisenberg limit, in phase contrast polarization microscopy[11,12], magnetic field sensing[13] and solution concentration measurement[14].

However, up to now almost all experimental results in this sense consisted in proof of principle demonstrations[6], because of the difficulties in generating high photon number entangled states and a high photon flux (comparable with the one used in classical schemes) and protecting them from decoherence up to the detection, mainly limiting the optical and detection losses[15].

Some remarkable results have been obtained thanks to the recent progresses in the generation of highly non-classical single mode[16] and few modes squeezed states[17]: they have been successfully implemented in gravitational wave interferometry[2], for particle tracking in biological environment[18,19], and to some extent for beam displacement measurement[20,21] and optical magnetometry[22].

Most quantum enhanced imaging and sensing protocols have been obtained exploiting single or few spatial modes of the quantum probe beam, such as in the case of squeezing, and with single photon detection in the schemes based on two-photon correlated states. In both the cases, only one parameter of the system, namely a single point of the sample, can be probed in the single run. The reconstruction of the sample as a whole requires time consuming scanning and accumulation of many detection windows. Instead, the exploitation of a high number, namely thousands, of modes in the same run is the requirement for quantum enhanced wide field imaging. In practice the number of spatial details of a structure, which can be probed at the same time, are determined by the number of spatial modes enclosed in the illuminating field. Indeed, one of the challenges in quantum optics and quantum enhanced imaging, is to generate and efficiently detect highly non-classical features in a multi-mode regime[23-28]. A first proof of principle of a quantum enhanced imaging protocol exploiting this parallelism has been reported in Ref.[29], following the proposal of Ref.[30], even if the average enhancement and the poor spatial resolution was not sufficient for any practical purpose, for example in absorption microscopy, where the technique is naturally addressed.

In this letter we address this point by presenting the realization of a sub-shot-noise (SSN) microscope exploiting thousands of spatial modes, detected independently by the same number of pixels of a charge-coupled-device (CCD) camera operated in the linear (non- amplified) regime. Thousands of photons per pixel are detected in the exposure time of the single shot. Therefore, the microscope operates in a wide-field regime (no scanning is required), suitable for dynamic imaging. It is based on the non-classical and spatially multimode correlations of squeezed vacuum, naturally generated by a traveling wave parametric amplifier both in low and in high gain regime[24,26,30-32]. The noise of the image, formed by the probe beam interacting with the sample, is locally reduced by subtracting pixel-by-pixel the correlated noise pattern measured on the other beam (reference)[30]. Moreover, we introduce the concept of quantum enhanced median filter[33]: quantum noise reduction at different spatial scales can be naturally combined with the statistical noise smoothing used in very standard image processing algorithm, with impressive overall enhancement in the object recognition.

For the first time we reach a significant improvement of the sensitivity with respect to any classical absorption microscopy system at the same illumination level. Our present results completely outperform the previous proof of principle demonstrations[29,34], drastically improving the resolution by a factor 10-100 (depending on the sensitivity level), both in terms of pixels count and in terms of the size of the imaged details of the sample.

Wide field microscopy is the simplest, fastest, less expensive and oldest imaging modality used, for example, for live-cell imaging. It has the advantage of requiring the lowest photon dose, especially for absorbed light imaging. It is recognized nowadays that the lowest photon dose that achieves a measurable metric for the experimental question should be used. For instance, this is paramount to ensure that the cellular processes being investigated are not shifted to an alternate pathway due to environmental stress[35]. Indeed, the results presented here can have immediate application in many field starting from biology and biochemistry. Furthermore, the comparison with the reference beam can be used to provide the absolute value of the absorption, providing the possibility of a quantitative analysis of the properties related to it. With small modifications (essentially in the data processing), our technique can also be the basis for getting enhancement sensitivity in schemes with different goals: for example ghost imaging[36-41], detection and imaging against environment or electronic noise background[42,43] and accurate characterization of retina rod-cells response to single photon stimulation[44]. Finally, this new capability finds application in quantum radiometry as well, for example to the absolute calibration of detectors with spatial resolution, as demonstrated in[45-48].

## MATERIALS AND METHODS

In absorption wide-field imaging, like in standard microscopy, a probe illuminates the sample all at once and its transmitted pattern, is imaged to the detector, typically the pixels array of a camera. The intensity measured by each pixel, $N_\alpha$, here expressed in number of photons, has a mean expectation value $\langle N_\alpha \rangle = (1-\alpha)\langle N \rangle$, $\langle N \rangle$ being the mean photon of the beam and $\alpha$ the absorption coefficient. The photon noise of the measurement can be obtained by modeling the absorption as the action of a beam splitter of transmittance $1-\alpha$ on the beam with initial variance $\langle \Delta^2 N \rangle$ [49]. It results in the form $\langle \Delta^2 N_\alpha \rangle = [(1-\alpha)^2(F-1) + 1-\alpha]\langle N \rangle$, where $F = \langle \Delta^2 N \rangle / \langle N \rangle$ is the Fano factor in absence of the sample. The value $F=1$ establishes a bound between classical and non-classical photon statistics. In particular, $F$ is lower bounded by the unity for classical states, while specific non-classical states may have sub-Poissonian photon statistics, i.e. $0 \leq F < 1$. The uncertainty on the absorption estimation in the direct (DR) imaging scheme is therefore

$$\Delta \alpha_{dr} = \frac{\sqrt{\langle \Delta^2 N_\alpha \rangle}}{|\partial_\alpha \langle N_\alpha \rangle|} = \sqrt{\frac{(1-\alpha)^2(F-1)+1-\alpha}{\langle N \rangle}} \tag{1}$$

The limit of the sensitivity for a classical probe ($F = 1$) is $\Delta \alpha = \sqrt{(1-\alpha)/\langle N \rangle}$, representing, for small absorption, the shot-noise limit with the typical scaling of the inverse square root of the number of photons. However, by inspecting Eq.(1), it is clear that non-classical optical fields with Fano factor smaller than one allows beating the shot noise limit. We note that the Fano factor, appearing in Eq.(1), is usually deteriorated, with respect to its value $F_0$ of the unperturbed (pure) state, by effect of the optical losses, including detector quantum efficiency. In particular, one gets $F = \eta F_0 + 1 - \eta$ where we have defined the overall detection probability $0 \leq \eta \leq 1$. Thus, the non-classical behaviour, in terms of noise reduction, is lower bounded by $F_{loss} = 1 - \eta$. It must be emphasized that splitting a single mode beam in $n$ pixels leads to a detection probability of the order of $\eta \leq 1/n$ for each of them, ruling out the possibility of using single mode for sub-shot-noise wide-field imaging for any reasonable number of pixels. Thus, it is evident the necessity of having many non-classical spatial modes, each one addressing a single pixel with limited losses. Even if sub-poissonian light beams have been obtained as a single or few modes, it is completely not obvious how to generate a beam with a high number of sub-Poissonian modes and how to detect them simultaneously. On the other side, it is relatively simple to produce a pair of beams, which are (individually) spatially incoherent, but locally correlated at the quantum level, by means of traveling wave parametric amplifier in the spontaneous regime. Even if the fluctuations of single spatial mode in one beam are super-poissonian, these fluctuations are perfectly replicated in the correlated mode of the second beam, because of photon number entanglement. This correlation is verified for all the wide range of localized transverse spatial modes. The degree of correlation and its non-classical features can be quantified by the noise reduction factor (NRF) $\sigma = \langle \Delta^2(N_1 - N_2) \rangle / \langle N_1 + N_2 \rangle$ [23,24,26,30-32,49-52], measured for a pair of pixels collecting correlated spatial modes. The NRF represents indeed the equivalent of the Fano factor for bipartite state, where the shot noise level is now $\langle N_1 + N_2 \rangle = 2\langle N \rangle$. While for classical beams the NRF is lower bounded by 1, quantum correlation can lead to $0 \leq \sigma < 1$. In particular, in presence of losses, $\sigma = 1 - \eta$.

While twin beams in a single spatial mode have been used for demonstrating sub-shot-noise absorption measurement in a double-beam scheme[53], Parametric Down-Conversion (PDC) multi-mode quantum correlations can be used for wide-field sub-shot-noise imaging[30].

Basically the object is placed in one beam and the second beam is used as a reference. Note that the double-beam (or double-path) approach is commonly used in imaging and spectroscopy where faint absorptions are involved, because it allows canceling out classical (super-poissonian) noise and provides a direct estimation of the absolute transmittance (absorption) by the instantaneous comparison with the unperturbed reference beam. We will consider the intensity difference between two correlated pixels of the two beams, whose expectation value is $\langle N_1 - N_{2,\alpha} \rangle = \alpha \langle N \rangle$ (we have assumed balanced beams, $\langle N_1 \rangle =$ $\langle N_2 \rangle = \langle N \rangle$). In this context the noise can be expressed in terms of the noise reduction factor in absence of the sample, in the form $\langle \Delta^2(N_1 - N_{2,\alpha}) \rangle = [\alpha^2(F-1) + \alpha + 2\sigma(1-\alpha)]\langle N \rangle$ (see Ref[30]).

Therefore, the absorption uncertainty in the differential (DF), both classical and quantum, imaging scheme is:

$$\Delta\alpha_{df} = \frac{\sqrt{\langle \Delta^2(N_1 - N_{2,\alpha}) \rangle}}{|\partial_\alpha \langle N_{-,\alpha} \rangle|} = \sqrt{\frac{\alpha^2(F-1) + \alpha + 2\sigma(1-\alpha)}{\langle N \rangle}} \qquad (2)$$

The performance of the classical differential (DC) imaging scheme is derived from Eq. (2) substituting $\sigma = 1$. In the situation of interest the absorption is so small that the terms $\alpha^2(F-1)$ are negligible even in presence of classical super-Poissonian noise of the source ($F > 1$) and the uncertainty in the differential classical imaging scheme becomes $\Delta\alpha_{dc} = \sqrt{(2-\alpha)/\langle N \rangle}$, just a factor 2 larger than shot-noise-limited direct imaging. Under the same condition, the quantum enhancement provided by sub-shot-noise (SSN) correlations with $\sigma < 1$, is quantified also in terms of Signal to Noise Ratio $\text{SNR} = \alpha/\Delta\alpha$ by using Eq. (1) and (2), as

$$\frac{\Delta\alpha_{ssn}}{\Delta\alpha_{dc}} = \frac{SNR_{dc}}{SNR_{ssn}} = \sqrt{\frac{\alpha + 2\sigma(1-\alpha)}{2-\alpha}} \approx \sqrt{\sigma} \qquad (3)$$

$$\frac{\Delta\alpha_{ssn}}{\Delta\alpha_{dr}} = \frac{SNR_{dr}}{SNR_{ssn}} = \sqrt{\frac{\alpha + 2\sigma(1-\alpha)}{1-\alpha}} \approx \sqrt{2\sigma}.$$

The advantage with respect to the differential classical scheme appears when $\sigma < 1$, while a more strict condition, $\sigma < 1/2$, is needed to beat the direct (shot-noise-limited) one.

The setup for the sub-shot-noise microscope scheme is sketched in Fig.1(a). A CW laser-beam (100 mW at $\lambda_p = 405$ nm) pumps a $1 cm$ Type-II Beta-Barium-Borate (BBO) nonlinear crystal, where two correlated beams are generated. The far field of the emission, where spatial correlation occurs, is realized at the focal plane of a lens with $f_{FF} = 1\ cm$ focal length. Then, the far field plane is imaged (magnification factor $M = 7.8$) to the detection plane by means of a second lens system with $f_{IM} = 1.6\ cm$. The detector is a charge-coupled-device (CCD) camera Princeton Inst. Pixis 400BR Excelon, operating in linear mode (no electro-multiplication gain), with high quantum efficiency ($> 95\%$ at $810\ nm$), 100% fill factor and low noise (read noise is few $e^-/(pix \cdot frame)$). The size of physical pixels of the camera is $13 \mu m$, but here we group them by a $3 \times 3$ hardware binning. Hereinafter, if not explicitly indicated, the single elementary pixel is intended having linear size of $39 \mu$m. This basically allows reducing enough the effect of the read noise for our purposes. A spectral selection is performed by two identical interferential filters ($800 \pm 20 nm$, with transmission of 99%), one just after the crystal and the other mounted on the camera. A test sample, with absorption $\alpha = 1\%$, representing the Greek letter "Φ" (size $300 \mu m \times 400 \mu m$) is realized by a few nanometers thick titanium deposition on a coated glass-slide.

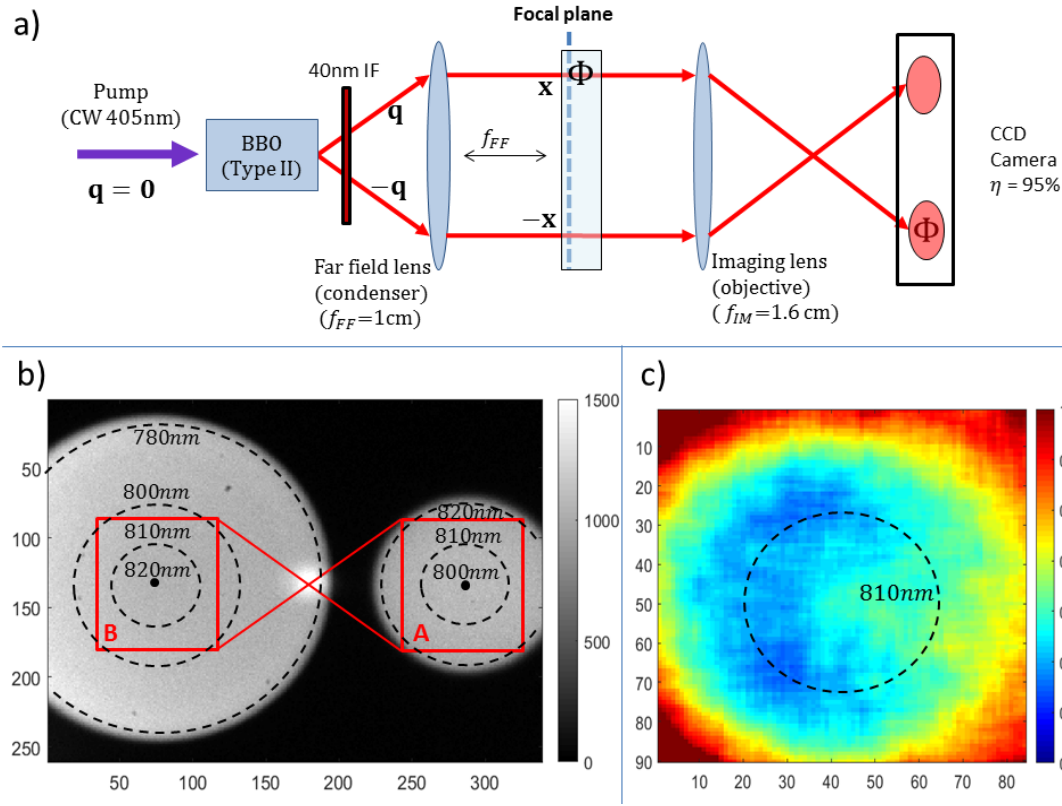

FIG. 1: Experimental Setup. a) Schematic of the experiment (for description see text). b) Image of the single shot acquired by the CCD camera in 100ms. The pixel size is 39μm, corresponding to 5μm resolution in the focal (object) plane. c) 2-D map of the NRF for 5μm resolution in the focal (object) plane. The axes are reporting the pixels number.

When inserted, the letter intercepts one beam at the focal plane of the far field lens, while the slide extents to the second beam (in the far field the centers of the two beams are separated by $1.0\ mm$). It is important to notice that the acquisition time of a single shot is typically $10^{11}$ times the coherence time of the PDC process (of the order of $10^{-12}s$). Considering that the number of photons per pixel is about $10^3$ the occupation of the single temporal mode is $10^{-8} photons/mode$. Therefore, the statistic of the noise is expected to be Poissonian, i.e. shot-noise limited.

In the crystal, a photon of the pump beam is converted in a pair of photons with lower frequency, fulfilling the energy and momentum conservation (phase matching-condition). In particular if the pump is approximated by a plane wave with transverse momentum $\mathbf{q} = 0$, the two photons of a pair must fulfill $\mathbf{q}_1 + \mathbf{q}_2 = 0$, thus being emitted with opposite transverse momenta. The lens maps momentum $\mathbf{q}$ at wavelength $\lambda$ into the point $\mathbf{x} = \lambda f \mathbf{q}/2\pi$ at its focal plane, in particular leading degenerate photons (having $\lambda_1 = \lambda_2 = \lambda_d = 2\lambda_p = 810\ nm$) to be found in symmetric position $\mathbf{x}_1 = -\mathbf{x}_2$ with respect to the pump direction Fig. 1(a). It is expected that two symmetric pixels of the camera always detect the same number of photons. Moreover, even if the emission is broadband both in frequency and momentum, phase matching conditions establish a relation between the wavelength and the photon direction, specifically the modulus of the transverse momentum, so that photons are emitted in concentric cones each corresponding to a certain wavelength. In Type-II PDC, the correlated photons have orthogonal polarizations and emission cones have different centers. The intensity distribution, detected in a 100ms shot is shown in Fig.1(b). The dotted circumferences approximately represent the wavelength distribution in the two orthogonally polarized beams. By selecting two symmetric regions *A* and *B* around degeneracy (represented by red squares in the picture), one expects each pixel of *A* to be non-classically correlated with a corresponding symmetric pixel in *B*. In practice there are two important limiting factors to this imaging system, which must be taken into account. One basically limits the field of view and the other lower bounds the spatial scale at which the noise subtraction can be efficiently performed.

-The trade-off between field of view and NRF: Around the degeneracy wavelength for correlated photons hold $\lambda_1 = \lambda_d + \Delta\lambda$ and $\lambda_2 = \lambda_d - \Delta\lambda$ and the point-to-point correlation in the far field becomes $\mathbf{x}_1 + \mathbf{x}_2 \cong \frac{2\Delta\lambda}{\lambda_d} \mathbf{x}_1$. Therefore, the center of symmetry of the correlations is $\mathbf{x}=0$ only at the degenerate wavelength. As long as one moves from the degeneracy, the center of symmetry shifts proportionally to $\frac{\Delta\lambda}{\lambda_d}$. Once the pixel grid has been positioned to be symmetric with respect to $\mathbf{x}=0$ only a relatively small spectral bandwidth around degeneracy can be tolerated (in our case about $40nm$). This reflects on the available angular bandwidth, see Fig.1(b), in our case corresponding to about $500\mu m \times 500\mu m$ field of view in the focal plane. Fig.1(c) shows a map of the NRF obtained by subtracting locally pixel-by-pixel the two regions *A* and *B*. Along the degenerate ring we have the best NRF, while moving far from it the correlation slightly decreases because the corresponding pixels of the two regions are no longer perfectly intercepting correlated directions. Moreover, on the right hand part we observe a further increasing of the NRF. Based on our experience, this can be ascribed to a small aberration of the optical system, which we could reduce but not completely suppress. Anyway, this technical issue can be solved by a careful analysis and realization of an ad-hoc optical system.

- The trade-off between spatial resolution and NRF: In practice, rather than a plane

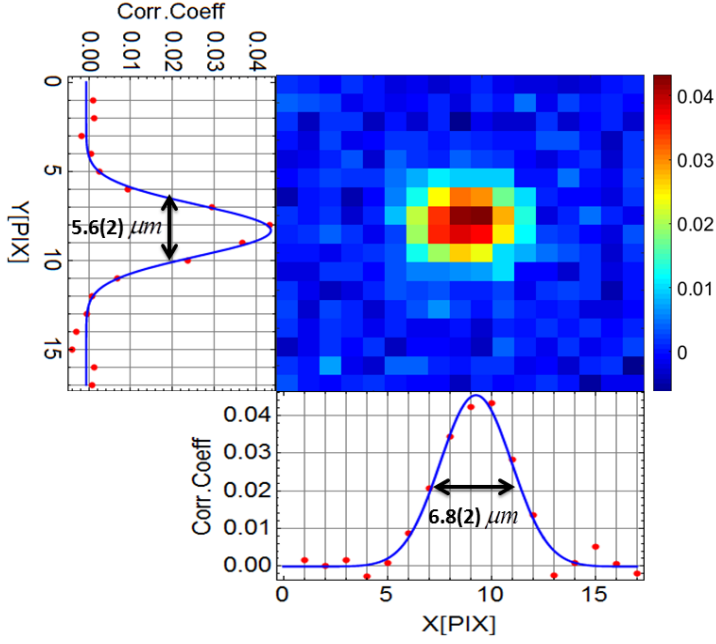

FIG. 2: Spatial cross correlation function. The 2-D map represents the value of the correlation coefficient between two regions of 40 x 40 pixels, chosen approximatively in symmetric position at the detection plane. Here we exploited the full resolution of the camera (physical pixel of 13μm) which corresponds to 1.7μm in the object plane. The peak represents the position in which the regions are well correlated pixel-by pixel. Shifting one of the region in the pixel grid of more than the spatial jitter of correlated photon makes the correlation coefficient dropping to zero. The vertical and horizontal sections are shown in the left-hand-side and bottom graphs respectively, with their fitting Gaussian functions and the indication of the FWHMs.

wave, the pump is a Gaussian beam propagating along $z$ direction, with waist $w_p$ and thus with a transverse momentum distribution centered at $\mathbf{q}=0$ with bandwidth $\Delta q \sim 2/w_p$. Transverse momentum conservation $\mathbf{q}_1 + \mathbf{q}_2 = 0 \pm \Delta q$ leads to a less strict position correlation in the far field, $\mathbf{x}_1 + \mathbf{x}_2 = \pm \Delta x$, where the relative uncertainty on the photon position is $\Delta \mathbf{x} \equiv 2r \sim \lambda f/\pi w_p$. A measurement of this spatial uncertainty is provided in Fig. 2, representing the spatial cross correlation function of the noise patterns of two symmetric regions of $40 \times 40$ physical pixels. The cross section is Gaussian with two slightly different FWHM in horizontal and vertical axes estimated to be $2r_y = 5.6(0.2)\mu m$ and $2r_x = 6.8(0.2)\mu m$ respectively. It is clear that two symmetric pixels detect most of the correlated photons only if their size $L$ is larger than this uncertainty (see Ref[32]).

## RESULTS AND DISCUSSION

The NRF can be described by the function $\sigma = 1 - \eta_0 \cdot \eta_{coll}$ where we split the detection probability $\eta$ in a term $\eta_0$, representing the transmission-detection efficiency of the optical path, and in $0 < \eta_{coll} < 1$, representing the collection efficiency of correlated photons. The efficiency $\eta_{coll}$ is a monotonic increasing function of the ratio $L/2r$, which reaches the asymptotic value $\eta_{coll} \sim 1$ for $L \gg 2r$ (see Supplementary Information Section). This means that at different resolution scales, given by the size of pixels or, more in general, by the spatial scale in which the signal is integrated, the noise reduction factor and thus the SNR, according to Eq. (3), are different. Anyway, the spatial information at different scales can be recovered starting from a high resolution image where the pixels size $x_{pix}$ can be smaller than the correlation area and by averaging the signal in a groups of $d \times d$ so that $L = d \cdot x_{pix}$. This is particularly useful considering that it can be seen as a quantum enhanced version of the standard median filter used for noise correction in classical imaging. In the classical median filter the value of a pixel is replaced with the mean value of its $d \times d$ neighborhood, which allows reducing the noise simply by a statistical cancellation, of course at the expense of filter details smoothing. If the median filter is applied to the noise- subtracted image, the two effects, namely the statistical smoothing and the photon noise subtraction, combine together, allowing a net improvement of faint object recognition. Fig. 3 shows an example of the application of the median filter to a direct image, to a classical differential image and to a SSN image. The upper-right panel faithfully represents the shape of the object obtained by averaging over 300 shots at full resolution ($L = 5\ \mu m$). The image represents the absorption coefficient, assuming values around zero outside the "Φ" and value around $\alpha = 0.01$ inside the "Φ". The other panels present single shot images at different resolution scales . The SSN images are obtained by subtracting to the direct image the correlated noise. The DC images are simulated by subtracting an uncorrelated (but shot-noise-limited) noise pattern, for example a noise region shifted of more than $2r$ from the correlated position. Notice that the single shot image of the sample is completely hidden by the shot noise at the full resolution, $d = 1$. When the median filters are applied, $d > 1$, one can appreciate a clear emerging of the shape of the sample, especially for the SSN case. As discussed before, the advantage of the SSN image increases with the scale $d$ of the median filter. The DC image is the worst because it contains twice the shot noise. It is worth noting that the DC imaging is advantageous with respect to the DR imaging if classical super-Poissonian noise is present, in particular when $F > 2$.

The exposure time of the single shot in our setup is 100 ms and the read-out time of the pixel matrix is few hundreds of milliseconds. In few seconds it is possible to realize a sub- shot-noise movie (file attached) in which the sample is simply translated by a micrometer stage during the acquisition. Even in this case the edges of the moving object are easier to be followed in the SSN imaging box (right-hand side window) rather than in the DC imaging (central window) or in DR imaging box (left-hand side window). This demonstrates that our technique is suitable for dynamic imaging. With more powerful or pulsed pump laser and faster operating modes achievable by commercial cameras, rate of hundreds frame/s should be reasonably reached.

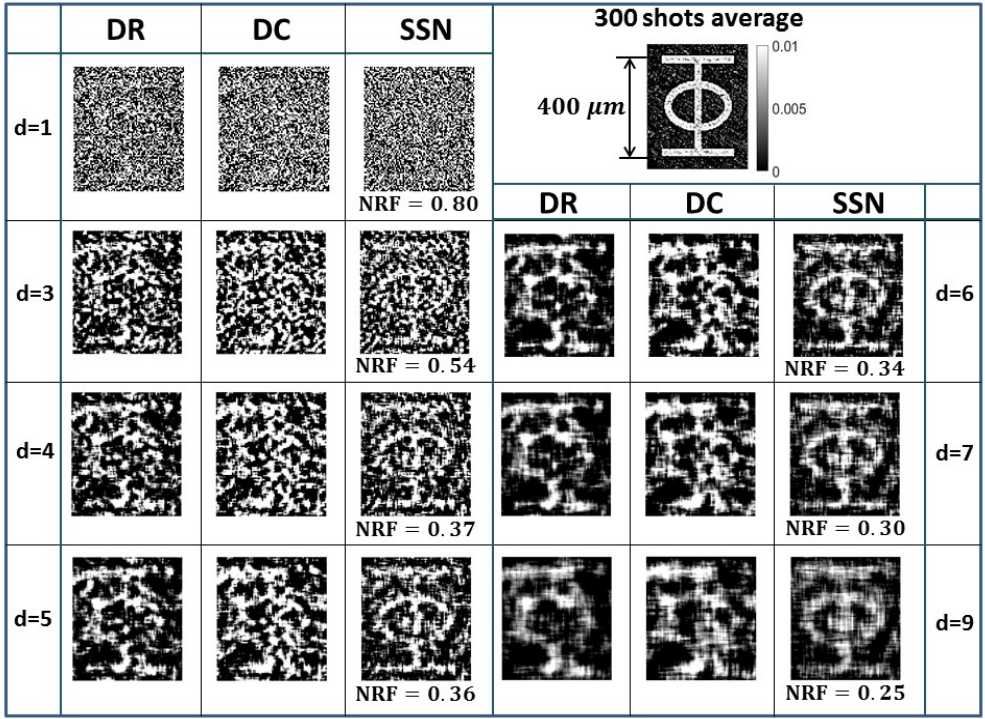

FIG. 3: Application of a median filter to a single shot image for different integration scales *d*. The direct (DR) image, the differential classical (DC), and the sub-shot-noise (SSN) one are compared in each panel for the same value of *d*. Upper-left panel is the image of the object after the average over 300 shots.

The statistical analysis of the SNR has been done by acquiring $\mathcal{N} = 300$ shots with and without the sample. The shots without the sample are used to estimate the NRF and Fano factor at different scales and also to eliminate static intensity and efficiency gradients by a standard flat field algorithm. Each shot with the sample contains a DR shot-noise limited image and the correlated noise. The NRF of the $n$-th shot is evaluated experimentally by spatial statistics over the ensemble of the correlated pixel pairs belonging to the region *A* and *B*, namely $\sigma(n) = V_x[N_A^{(n)}(\mathbf{x}) - N_B^{(n)}(\mathbf{-x})]/E_x[N_A^{(n)} + N_B^{(n)}]$ where $E_x[N(\mathbf{x})] = (1/\mathcal{M})\sum_x N(\mathbf{x})$ is the mean value of the $\mathcal{M}$ pixel of the region, and $V_x[N(\mathbf{x})] = E_x[N^2(\mathbf{x})] - E_x[N(\mathbf{x})]^2$ is its variance. The average NRF over the 300 values is reported in the graph in Fig. 4 for different resolution scales in the object plane (see also Supplementary Information Section). In particular, we note that already for resolution of 5 $\mu m$, comparable with the correlation FWHM of the spatial correlation function, the system reaches a NRF of $\sigma = 0.8$, allowing to beat DC imaging (see the first of Eq.s (3)) in a matrix of more than 8000 pixels. For 15 $\mu m$ resolution the NRF is below 0.5, which is the threshold for beating the performance of the DR, sub-shot-noise limited, imaging (see the second of Eq.s (3)). Similarly, also the Fano factor is evaluated and reported in Fig. 4, demonstrating the Poissonian character of the light statistics.

The SNR is estimated over a stripe of pixels of the images after the application of the median filter (in particular we considered a vertical stripe inside the main axis of the letter "Φ"). Let us label $\alpha^{(n)}(\mathbf{x})$ the absorption value of the pixel in position $\mathbf{x}$ of the $n$-th shot. First, the SNR for each position evaluated as $SNR(\mathbf{x}) = E_n[\alpha(\mathbf{x})]/V_n[\alpha(\mathbf{x})]^{1/2}$ where $E_n[\alpha(\mathbf{x})] = (1/\mathcal{N}) \sum_{n=1}^{\mathcal{N}} \alpha^{(n)}(\mathbf{x})$ is the experimental temporal average of the absorption and $V_n[\alpha(\mathbf{x})]$ is its variance. Then, the spatial average of the $SNR(\mathbf{x})$, for $\mathbf{x}$ belonging to the vertical stripe, is evaluated. The experimental results, showing the advantage of the quantum noise subtraction, are reported in Fig. 4. The data are compared with the theoretical prediction obtained by substituting the estimated NRF in the theoretical expression in Eq. 3. One can note that the SNR improvement of the SSN imaging is slightly higher than expected, with respect to both the DR and DC imaging. This is explained from the fact that the SNR is evaluated on a vertical stripe, which lies close to the degenerate wavelength region, where the NRF is slightly lower than the average (see Fig.1(c) and the related discussion).

**CONCLUSIONS**

In summary, we have realized the first sub shot noise wide field microscope, demonstrating a noise reduction of 20% below the shot noise for each resolution cell (pixel) of $5\ \mu m$ in a matrix of about 8000 pixels. This is sufficient for wide-field imaging of complex structures. Increasing the spatial scale of the details, the noise reduction improves accordingly, e.g. for resolution of $25\ \mu m$ is 62% below shot noise and for $50\ \mu m$ is 72%, on average, below shot noise level. At this last scale, for example, it almost doubles the SNR of the classical differential technique for the same illumination level, while the sensitivity improvement with respect to classical direct imaging (obtained by a single, shot noise limited beam) is around 30%. Equivalently it allows one to maintain the same SNR reducing either the exposure time or the illumination level of almost four times compared with the DC imaging and almost twice compared to DR imaging. In general in order to have the SNR of DC (DR) the photon number of quantum beam can be reduced by a factor (2) $\sigma$. We have shown that the reduction of the quantum noise at different scales is perfectly compatible with, and actually improves, the standard noise reduction techniques based on a posteriori elaboration of the image taken at full resolution. We demonstrated this important point with a sample made by a ultra-thin metallic depositions on glass slide. These performances represent a breakthrough, filling the gap between the prof of principle of quantum enhanced scheme and a system suitable for potential application[6,56].

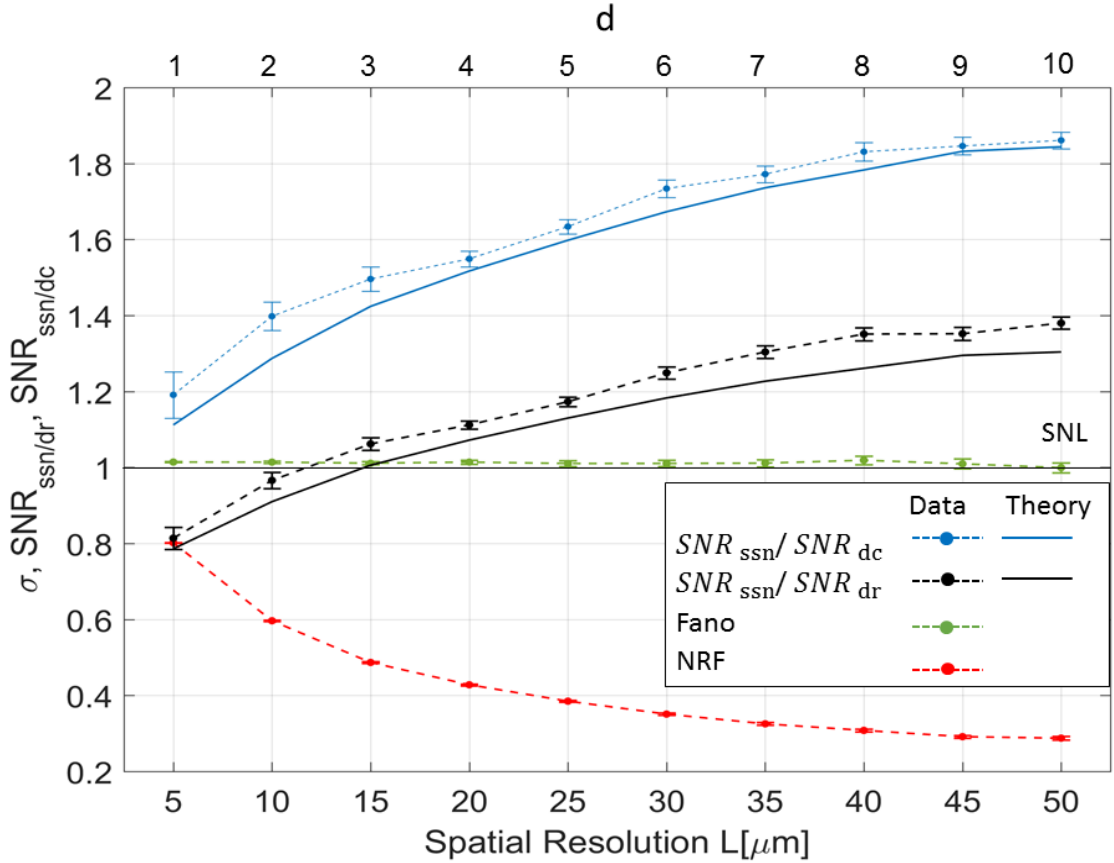

FIG. 4: Experimental noise reduction factor (NRF) and signal-to-noise ratio (SNR) in function of the resolution in the focal (object) plane *L* (or equivalently in function of the binning *d* of the median filter, upper scale). Red dots represents average of the NRF in a region of the same size as the "Φ" object, i.e. $400 \times 300 \mu m$. The black dots are the SNR of the sub-shot-noise images normalized to the one of the direct images. For *L* ≥ 15μm there is the advantage of the quantum protocol. Analogously, the blue series shows that advantage of the sub-shot-noise imaging with respect to the differential classical imaging is present at any spatial resolution and reaches values of more than 80%. Solid lines correspond to the quantum enhancement predicted by Eq.s (3), when the estimated values of the NRF are considered

We believe that our technique has the potentiality for a wide-spread use in absorption microscopy. The spatial resolution can be further improved (nothing prevents to get close to the Abbe limit), and the range of applicability can be extended engineering highly non- classical bright squeezed vacuum sources in pulsed regime[52].

**ACKNOWLEDGMENTS**

This work was supported by the MIUR Project "Premiale P5". We thank Matteo Fretto for the micro-fabrication of the object used in the measurement and I. P. Degiovanni for fruitful discussions and advises.

**AUTHOR CONTRIBUTIONS**

IRB, AM, MG conceived the idea of the experiment, that was discussed and designed with inputs by all authors. NS, IRB, AM realized the experimental setup and collected the data in INRIM quantum optics labs (coordinated by MG) . All authors discussed the results and contributed to the writing of the paper.

**CONFLICT OF INTEREST**

# SUPPLEMENTARY INFORMATION

We briefly present a model describing the non-classical characteristic of the system in terms of noise reduction factor addressed to realize sub-shot noise (SSN) imaging. For any pair of detectors $D_1$ and $D_2$ (e.g. two pixels of the camera), collecting $N_1$ and $N_2$ photons respectively, the noise reduction factor (NRF) [23,24,26,30-32,49-52] is defined as

$$\sigma = \frac{\langle \Delta^2(N_1 - N_2) \rangle}{\langle N_1 \rangle + \langle N_2 \rangle} \tag{1}$$

Here the numerator is the variance in photon number difference expressed as

$$\langle \Delta^2(N_1 - N_2) \rangle = \langle \Delta^2 N_1 \rangle + \langle \Delta^2 N_2 \rangle - 2\langle \Delta N_1 \Delta N_2 \rangle \tag{2}$$

where the last term represents the covariance. The detectors $D_1$ and $D_2$, with quantum efficiency $\eta_1$ and $\eta_2$ respectively, measure the photon number in the far field of a spatial multimode twin beam (TWB) with mean photons per spatio-temporal mode µ. According to the main text description, the correlated photons can be found in symmetric positions with spatial uncertainty $\Delta x \equiv 2r = \lambda f/\pi w_p$ where $w_p$ is the pump width. $\Delta x$ represents the size of the correlated spatial modes at the detection plane in the far field. Spatial uncertainty in both transverse directions, $x$ and $y$, are considered equal in our model. Referring to Fig. S1, we consider two detectors, $D_1$ and $D_2$, of size $L \times L$, aiming at intercepting correlated modes, which are symmetric with respect to a point (red dot in Fig. S1). Certain number of spatially correlated modes $M_c$ (marked in color red), are efficiently detected inside the detection area. Due to possible misalignment $\delta$, certain number $M_u$ of collected modes are not pair wise correlated (colored in blue). Modes $M_b$, in the border of the detection area (marked in color green), are collected with $\beta = 50\%$ efficiency on average. Taking into account the different contribution of the various types of spatial modes to the multi-thermal photon number statistics in each beam (we do not consider temporal modes here), one has[55]:

$$\langle N_1 \rangle = (\beta M_b + M_c + M_u)\eta_1 \mu \tag{3}$$
$$\langle N_2 \rangle = (\beta M_b + M_c + M_u)\eta_2 \mu \tag{4}$$
$$\langle \Delta^2 N_1 \rangle = \eta_1^2 \mu^2 (\beta^2 M_b + M_c + M_u) + \eta_1 \mu(\beta M_b + M_c + M_u) \tag{5}$$
$$\langle \Delta^2 N_2 \rangle = \eta_2^2 \mu^2 (\beta^2 M_b + M_c + M_u) + \eta_2 \mu(\beta M_b + M_c + M_u) \tag{6}$$

For PDC the covariance between two detectors only involves correlated modes and it is expressed in the form[39]

$$\langle \Delta N_1 \Delta N_2 \rangle = \eta_1 \eta_2 \mu(\mu + 1)(\beta^2 M_b + M_c) \tag{7}$$

For the ideal condition, when the correlated modes are the total number of modes, the above expressions simplify to the usual form of multi thermal statistics: $\langle N_j \rangle = M_c \eta_j \mu$, $\langle \Delta^2 N_j \rangle = M_c \eta_j \mu (1 + \eta_j \mu)$, where $j = 1,2$. In the calculation of the NRF it is useful to formally balance the mean photon counts in the two channels, which can have different detection efficiencies, by introducing the factor $\gamma \geq 1$ so that $\eta_1 = \gamma \eta_2 = \eta_0$ (we consider $\eta_1 > \eta_2$) and by the substitution $N_2 \to \gamma N_2$ in Eq.s (1-2). Then, using Eq.s (3-7) in the Eq.s (1-2) after the previous substitution leads to the expression

$$\sigma = \frac{\gamma+1}{2} - \eta_0 * \eta_{coll}, \tag{8}$$

where $\eta_{coll} = \frac{\beta^2 M_b + M_c - \mu M_u}{\beta M_b + M_c + M_u}$ can be interpreted as the collection efficiency of the correlated photons. For the ideal case when the modes are perfectly correlated, i.e. $M_u = M_b = 0$ with no unbalancing, i.e. $\gamma = 1$, one recovers the usual expression for NRF, i.e. $\sigma = 1 - \eta_0$. Under the geometrical conditions $L > r$ and $\delta \ll L$, different types of mode are related to the measurable parameters as $M_u = 2L\delta/\pi r^2$, $M_c = [(L-2r)^2 - 2L\delta]/\pi r^2$, and $M_b = 2L/r$ (see Fig.S1), where $r$ is the coherence radius at the detection plane. By introducing the dimensionless parameters $X = L/2r$ and $D = \delta/2r$, the collection efficiency becomes

$$\eta_{coll} = \frac{X(\pi\beta^2 - 2D(\mu+1) - 2) + X^2 + 1}{X^2 + (\pi\beta - 2)X + 1} \tag{9}$$

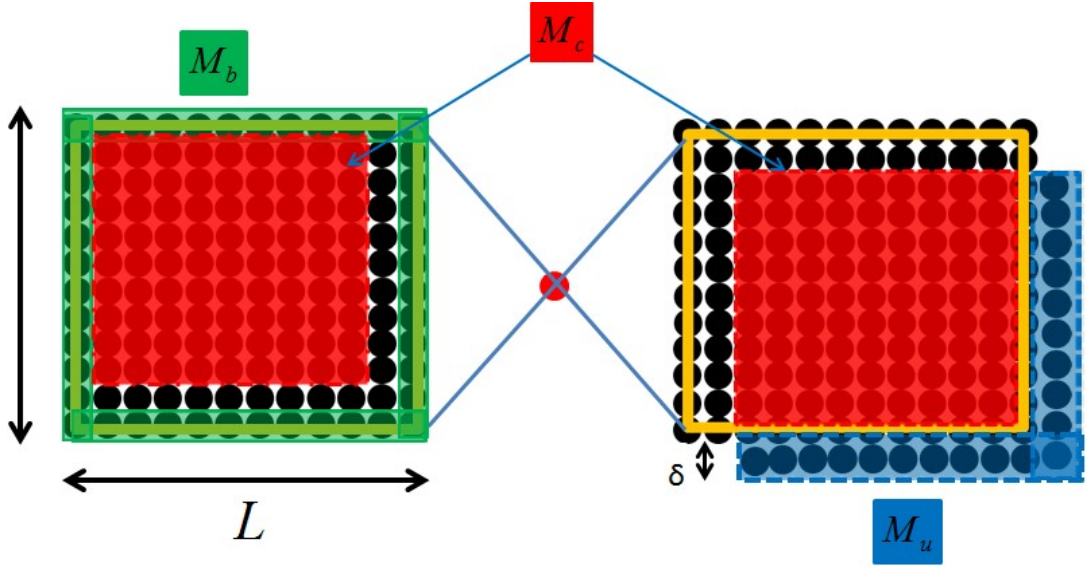

FIG. S1: Schematic of the detection area of size *L* and misalignment *δ*. The black filled circles represent ideally the TWB spatial modes at the far field plane which are pairwise correlated with respect to a center of symmetry indicated by the red circle in the center of the picture.

We work in low gain regime, with a mean photon number in a spatio-temporal mode $\mu \approx 10^{-8}$. This means that the muti-thermal statistics (in Eq. 3-6), converges to the Poissonian one and the dependence of $\eta_{coll}$ from μ can also be neglected in Eq. (9). Thus, in the limit $\mu \to 0$ NRF in Eq. (8) does not depend on the mean number of photons per pixel[32]. In the asymptotic limit $X \gg 1$, i.e. when the detection size is much larger than the correlation area, $\eta_{coll}$ and σ approach the unity and $1 - \eta_0$ respectively.

However, up to now we have not considered spurious noise. The measured NRF is affected by the independent noises coming mainly from the electronic read noise per pixel $\Delta^{(el)}$ of the CCD and the from the stray light $N^{(st)}$ (fluorescence of the laser pump in BBO crystal and in the interference filter). In particular, the total photon number $N_j^{(tot)}$ measured by each detectors $D_j$ is the sum of contributions from TWB and stray light, i.e. $N_j^{(tot)} = N_j^{(TWB)} + N_j^{(st)}$ ($j = 1,2$). Hereinafter, for simplicity we consider $N_1^{(tot)} = N_2^{(tot)} = N^{(tot)}$. The total noise of each detector is $\Delta^2 N_j^{(tot)} = \Delta^2 N_j^{(TWB)} + \Delta^2 N_j^{(st)} + (\Delta^{(el)})^2$, where the last part corresponds to the electronic read-noise of the CCD. NRF follows from Eq.s (1-2) as

$$\sigma_{eff} = \sigma f^{(TWB)} + f^{(noise)} \qquad (10)$$

where $f^{(TWB)} = (N^{(tot)} - N^{(st)})/N^{(tot)} = N^{(TWB)}/N^{(tot)}$ is the fraction of TWB photons and $f^{(noise)} = [(\Delta^{(el)})^2 + N^{(st)}]/N^{(tot)}$ is the fraction of spurious noise of the system (assuming poissonian behavior of the stray light). Eq. (10) shows how the spurious noise increases the experimental value of the NRF, and how this effect can be mitigated by increasing the fraction of twin beam photons. In our set up, fraction of stray light cannot be reduced arbitrarily since we are not allowed to use very narrow spectral filters. We have reached $N^{(st)}/N^{(tot)} = 3.8\%$, a ratio which does not depend on pump power or acquisition time because both stray light, and PDC light grow linearly in our operative regime. However, according to Eq.(10), increasing the total number of photons $N^{(tot)}$ (by rising the pump power in the linear gain regime, or increasing the acquisition time) allows reducing the effect of the electronic noise which has been measures as $\Delta^{(el)} = 4.9 \ e^-/(pix \cdot frame)$ for our camera settings.

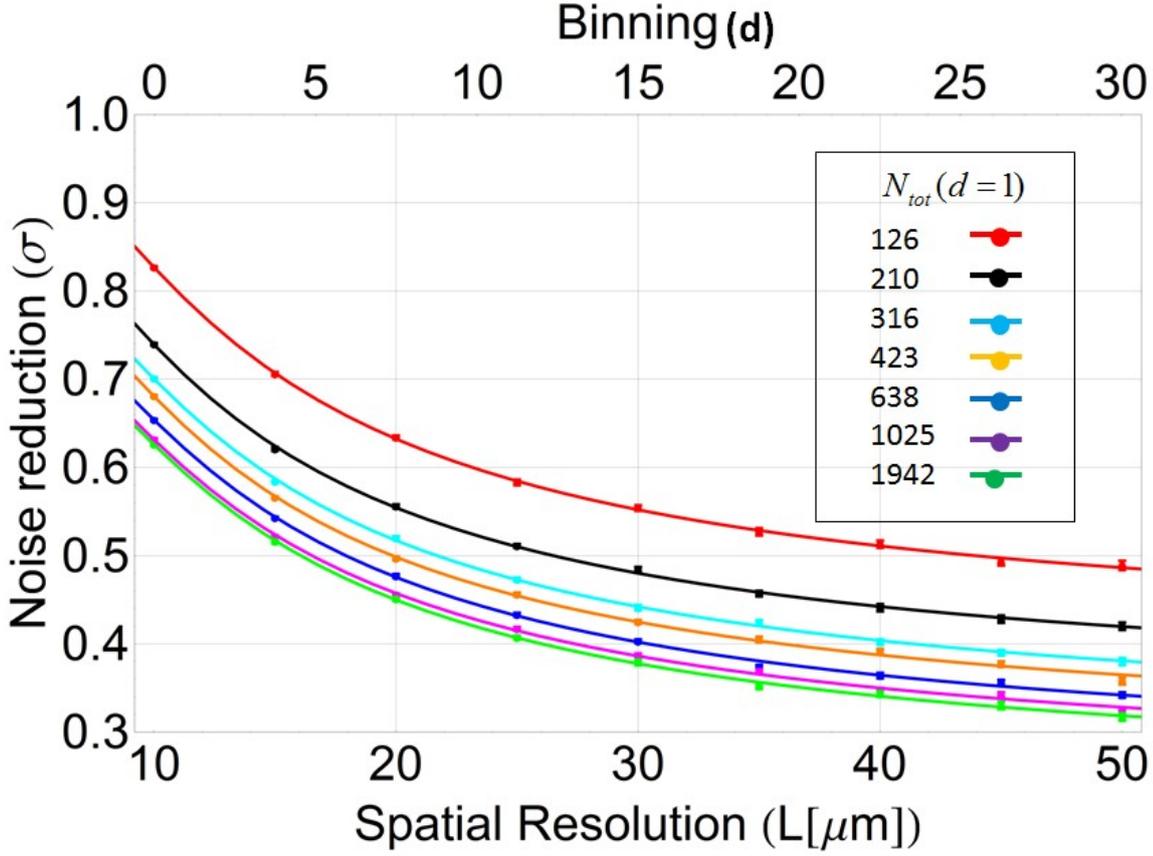

FIG. S2: Noise reduction vs. Spatial resolution. Equivalence in terms of binning is shown in top scale.

Fig. S2 shows the effective noise reduction factor $\sigma_{eff}$ as a function of the resolution, for different number of photon per pixel $N_j^{(tot)}$ of the initial (non-binned) image. The resolution, i.e. the effective pixel size $L$, is changed practically by a binning procedure, which consists of taking the sum of the photon numbers in groups of $d \times d$ pixels so that $L = d \cdot x_{pix}$, where $x_{pix}$ is pixel size of the initial image. For each shot, in a set of 300, the NRF is evaluated by measuring the spatial variance, normalized to the sum of the mean value, of the binned and subtracted regions. Then, the 300 values are averaged and are represented in the data points. The different curves correspond to different total mean photon number $N^{(tot)}$ are obtained by changing the pump power, always remaining in the linear regime of PDC (namely $\mu \leq 10^{-8}$). The experimental points are well fitted by the model Eq. (10). The coherence radius $r$, the efficiency $\eta_0$ and the misalignment $\delta$ are the free parameter of the fit. The different curves provide seven values of the free parameters, which are all consistent. The average values are $\eta_0 = 0.81(0.003)$, $r = 2.64(0.06)\mu m$ and $\delta = 0.63(0.1)$. Moreover, the value $2r$ obtained from the fit is close to the value of the FWHM of the cross correlation function estimated independently and quoted in Fig. 2 (in the main text).